# Lifetime of sub-THz coherent acoustic phonons in a GaAs-AlAs superlattice

A.A. Maznev[1], Felix Hofmann[1], Adam Jandl[2], Keivan Esfarjani[3], Mayank T. Bulsara[2], Eugene A. Fitzgerald[2], Gang Chen[3], and Keith A. Nelson[1]

[1]Department of Chemistry, Massachusetts Institute of Technology, Cambridge, MA 02139, USA.

[2]Department of Materials Science and Engineering, Massachusetts Institute of Technology, Cambridge, MA 02139, USA.

[3]Department of Mechanical Engineering, Massachusetts Institute of Technology, Cambridge, MA 02139, USA.

**Abstract**

We measure the lifetime of the zone-center 340 GHz longitudinal phonon mode in a GaAs-AlAs superlattice excited and probed with femtosecond laser pulses. By comparing measurements conducted at room temperature and liquid nitrogen temperature we separate the intrinsic (phonon-phonon scattering) and extrinsic contributions to phonon relaxation. The estimated room temperature intrinsic lifetime of 0.95 ns is compared to available calculations and experimental data for bulk GaAs. We conclude that ~0.3 THz phonons are in the transition zone between Akhiezer and Landau-Rumer regimes of phonon-phonon relaxation at room temperature.

Intense interest in phonons in nanostructured materials[1,2] arises both from the challenging physics involved and practical applications such as engineering of efficient thermoelectric materials[3] and thermal management of microelectronic devices.[4] Semiconductor superlattices (SLs) present excellent one-dimensional model systems and, consequently, both thermal phonon transport and coherent phonons in SLs attracted considerable attention.[1,5-9] Phonon relaxation time is one of the key parameters involved in both thermal transport and coherent phonon phenomena. Yet, phonon lifetimes at temperatures on the order of the Debye temperature are largely unexplored even for well-characterized bulk materials. Only recently, first-principles calculations of phonon lifetime for single crystals such as Si and GaAs have been accomplished within the three-phonon scattering model.[10-12] Experimental studies, on the other hand, lag behind, with room temperature phonon lifetime data available only at frequencies up to 100 GHz for Si[13] and up to 56 GHz for GaAs[14]. These data cannot be compared with the recent theoretical work[10-12] because at low frequencies the phonon dissipation is believed to be described by the Akhiezer relaxation model rather than three-phonon scattering (Landau-Rumer) model.[13-15] The transition between the two regimes, expected to occur somewhere in the sub-THz range,[13] has been neither measured nor calculated for any material.

For SLs, the situation is complicated by the fact that in addition to phonon-phonon scattering that typically dominates bulk phonon lifetime at room temperature, phonons in SLs are scattered by interface roughness and variations in layer thickness. Furthermore, measurements can be affected by inhomogeneous broadening due to non-uniformity of the SL period. The lifetime in an ideal SL structure due to phonon-phonon scattering will be referred to below as *intrinsic* lifetime.[9] It is highly desirable that this intrinsic lifetime be separated from other contributing factors that depend on the sample fabrication process. In the only hitherto reported measurement of phonon lifetime in an SL at room temperature[16] that yielded 250 ps at ~0.6 THz for a 8 nm-period GaAs-AlAs SL, the mechanism of the phonon dissipation was not clarified. A study of phonon lifetime in an acoustic cavity formed by two SL mirrors[17] concluded that the measured lifetime of ~100 ps at 1 THz was mostly extrinsic.

In this work, we study coherent sub-THz longitudinal phonons in a GaAs-AlAs SL generated and probed by femtosecond laser pulses. Based on the comparison of measurements made at room temperature and at 79 K we separate intrinsic and extrinsic contributions to phonon lifetime. The results will be compared with the available theoretical and experimental data for bulk GaAs.

The sample consisted of 219 periods of GaAs and AlAs layer pairs of 8 nm / 8 nm nominal thickness (the total period $d = 16$ nm) grown on an (100) GaAs substrate. It was fabricated by metal-organic chemical vapor deposition (MOCVD) at a temperature of 750 $^o$C. The precursors for the epitaxial growth were trimethyl-gallium, trimethyl-indium, and arsine. A 500 nm homoepitaxial layer of GaAs was initially deposited and was followed by the superlattice structure. The total thickness of the SL was 3.5 μm.

Figure 1(a) schematically shows the experimental arrangement. To generate and detect acoustic waves, we used an amplified Ti:Sapphire system (wavelength 784 nm, pulse duration 300 fs, repetition rate 250 kHz), whose output was split into excitation and variably delayed probe pulses. The excitation beam (pulse energy 57 nJ) was modulated by an acousto-optic modulator at 93 kHz frequency to facilitate lock-in detection and focused to a spot of 100 μm diameter (at 1/e intensity level) at the sample. The variably delayed probe pulse (energy 25 nJ) was focused to a 30 μm diameter spot at the center of the excitation spot. Upon reflection from the sample, the probe beam was directed to a photodiode, whose output was fed into a lock-in amplifier. The measurements were averaged over several hundred scans of the optical delay line, with a typical data collection time of ~20 hours. The sample was placed in a cryostat and the measurements were performed at room temperature (RT, 296 K) and at 79 K.

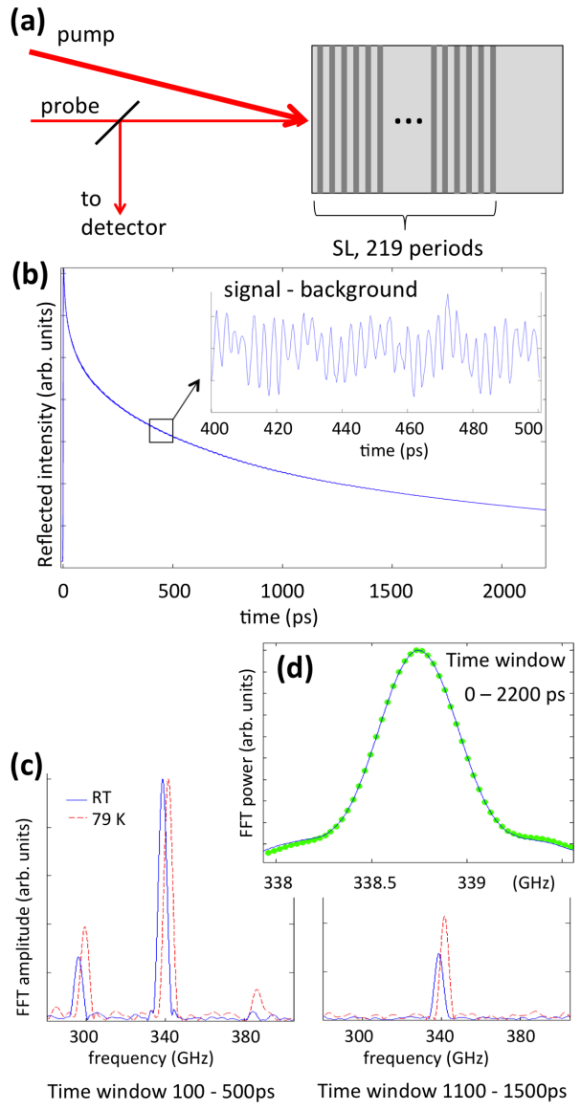

Fig. 1. (a) Schematic of the SL structure and the experimental arrangement (not to scale). (b) Measured signal waveform and acoustic oscillation trace obtained by subtracting the slowly varying background and multiplying by a factor of 50. (c) Fourier spectra of the acoustic oscillations showing the zone-center mode and "satellites" for a time window from 100ps to 500ps and from 1100ps to 1500ps. (d) Fourier spectrum (symbols) of the zone-center mode at RT in a large time window. Solid curve shows a fit to the Fourier-spectrum of a windowed exponentially-decaying sinusoid.

The mechanism of laser excitation and detection of coherent phonons in GaAs/AlAs structures is well known: absorption of excitation light in GaAs layers (AlAs being virtually transparent at the laser wavelength) generates an excited carrier population that leads to mechanical stress via deformation potential.[18,19] Acoustic waves generated by this transient stress modulate the effective optical constants of the SL and can be detected by monitoring the reflectivity of the structure with the probe laser pulse.

Figure 1(b) shows the signal waveform at RT dominated by the response to electronic excitation. Subtracting the slowly decaying background reveals acoustic oscillations, whose spectrum, shown by solid curve in Fig. 1(c), yields a low frequency Brillouin peak at 44 GHz (not shown) as well as a high frequency peak at 339 GHz flanked by two "satellites", at 296 GHz and a weaker one at 385 GHz. Also shown in Fig. 1(c) is the spectrum of the sample at 79 K (dashed curve), showing a small shifting of all peaks to higher frequencies. Comparison of the spectra for time windows from 100 to 500 ps and from 1100 to 1500 ps, shows a greater reduction in the height of the 339 GHz peak at RT than at 79 K, while the satellites disappear altogether.

In Figure 2 the measured frequencies are compared with calculated dispersion curves for longitudinal acoustic phonons in the SL.[5] We used the acoustic velocities values 4719 ms$^{-1}$ for GaAs and 5718 ms$^{-1}$ for AlAs taken from the literature[20,21] while the layer thicknesses were slightly adjusted to $d_{GaAs}$ = 7.38 nm and $d_{AlAs}$ = 7.84 nm to match the experimentally measured frequencies. The central 339 peak corresponds to the first symmetric zone-center mode with zero reduced wavenumber. This mode lies at the bottom of a small bandgap formed at the zone center whereas the zone-center mode at the top of the bandgap with antisymmetric strain pattern is not excited in our experiment. The satellites as well as the low-frequency peak are due to modes at wavenumber $q=2k$, where $k$ is the optical wavenumber in the medium. It is known[7] that the $q=2k$ modes appear to be short-lived compared to the zone-center mode due to their higher group velocity. This is clearly seen in Fig. 1(c).

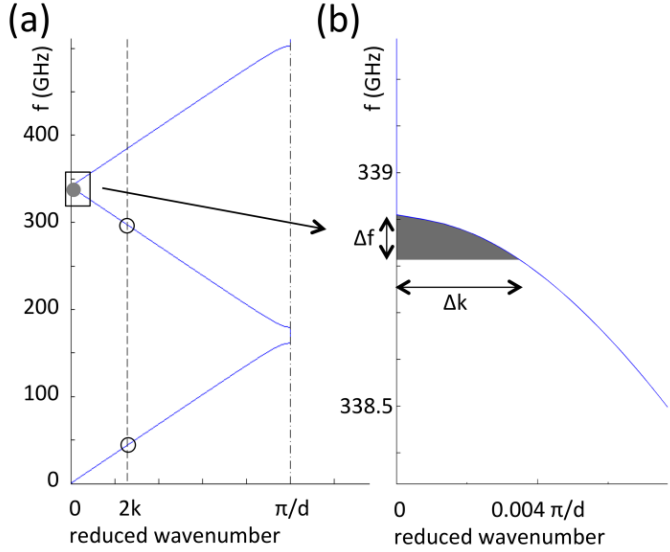

Fig. 2. (a) Measured frequencies (circles) and calculated dispersion curves for longitudinal phonons within the first mini-Brillouin zone of the SL at RT; the symmetric zone-center mode is shown as a filled circle, modes at the wavevector equal to twice the wavevector of light in the medium are shown as open circles; (b) Excited wavevector range $\Delta k$ and corresponding frequency spread $\Delta f$ for the zone-center mode.

Our attention here is focused on the long-lived zone-center mode. To measure its decay time, we performed Fourier transform in a sliding Hann window[22] with a width of 200 ps (FWHM) placed at 200 ps intervals. The evolution of the sliding Fourier transform amplitude at RT and at 79 K, normalized with respect to the amplitude at 260 ps, is shown in Fig. 3(a). It is clear that the amplitude decays more rapidly at RT. By fitting with an exponential function amplitude decay times of 800±20 ps and 1380±50 were determined for RT and 79 K, respectively. The phonon lifetime is commonly defined[14] as half of the 1/e amplitude decay time; hence the measured phonon lifetimes are $\tau_{RT}=400\pm10$ ps and $\tau_{79K}=690\pm25$ ps. The statistical error of the measurement was estimated by dividing each dataset, consisting of several hundred scans, into smaller sets of 20 delay stage scans, determining the decay time for each set and performing a standard statistical analysis. Alternatively, the lifetime can be determined by performing Fourier transform in a large window and fitting the spectrum with the Fourier transform of a windowed, exponentially decaying sinusoid, as shown in Fig. 1(d). This method yields similar results ($\tau_{RT}=393$ ps) to the sliding Fourier-transform analysis; however, the latter has an advantage of being a more direct model-independent method.

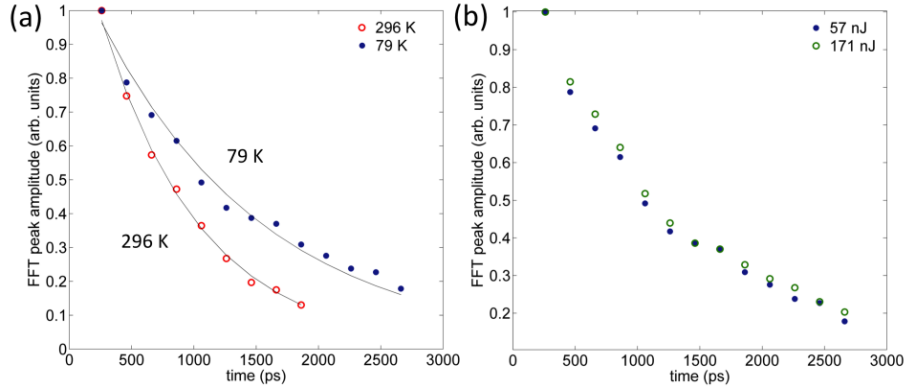

Fig. 3. (a) Time evolution of the sliding Fourier spectrum amplitude of the zone-center 340 GHz mode at temperatures 296 K and 79 K (symbols) fitted by exponential functions (solid curves). (b) Comparison of the Fourier spectrum amplitude time-evolution at 79 K for pump pulse energies 57 nJ and 171 nJ.

We verified that the decay time did not depend on the excitation energy. This is illustrated in Fig. 3(b) showing similar decay dynamics at two different excitation pulse energies, 57 nJ and 171 nJ, for measurements at 79 K. Consequently, interaction with the excited electrons which was suggested as a coherent phonon relaxation mechanism[23] does not play a significant role in our case.

Next, let us consider the effect of finite excitation depth. The zone center mode with the identically zero wavevector can only be generated by a spatially uniform excitation. The finite absorption depth $L$ leads to a spread in the wavevectors, $\Delta k=1/L$ which results in a finite width peak in the frequency domain, and, correspondingly, a finite lifetime. Normally one would estimate the frequency spread as $\Delta\omega=v_g\Delta k=v_g/L$, where $v_g=d\omega/dk$ is the group velocity. This expression has a simple physical meaning: $L/v_g$ is the time it takes for acoustic waves to "run away" from the surface. In our case, however, the excited modes cannot be characterized by a single group velocity value due to the parabolic shape of the dispersion curve $\omega(k)$ near the zone center. Therefore we estimate $\Delta\omega$ directly from the dispersion curve as shown in Fig. 2(b).

The absorption depth in bulk GaAs at 784 nm at RT is 701 nm, while AlAs is virtually transparent at this wavelength. In the effective medium approximation, the dielectric constant of a layered medium is given by a thickness-weighted average of the dielectric constants of the layers, which yields the absorption length $L$ equal to 1.447 µm, i.e. approximately twice of that in bulk GaAs. The actual absorption length is larger because of the quantum well effect that shifts the absorption edge in thin GaAs layers to shorter wavelengths.[24] The estimated lower bound for $L$ yields $\Delta k$ of 0.691 µm$^{-1}$, which leads, as shown in Fig. 2(b), to the upper bound for the frequency spread $\Delta f$ of 0.09 GHz, which is much smaller than the experimentally measured line width of ~0.5 GHz (see Fig. 1(d)). Thus the finite absorption depth should not have any effect on the observed lifetime at RT. At liquid nitrogen temperature, the absorption length in GaAs is greater than at room temperature,[25] which makes the effect even smaller.

The exclusion of the electron-phonon interaction and the "run-away" effect as contributors to measured phonon lifetime leaves three possible mechanisms: (i) phonon-phonon interaction; (ii) scattering by defects and interface roughness and (iii) inhomogeneous broadening due to possible non-uniformity of the SL period. Of these, only the first, "intrinsic" relaxation process is temperature-dependent. At temperatures higher that the Debye temperature, the intrinsic phonon lifetime scales linearly with temperature.[26] At low temperatures, the dependence becomes steeper due to rapid decrease in the population of high-frequency thermal phonons. According to first-principles thermal conductivity calculations for bulk GaAs,[27] decreasing the temperature from RT to 80 K results in about an order of magnitude increase in the phonon lifetime below 1 THz. In experiment,[14] a more than an order of magnitude increase in lifetime between 300 and 80 K was observed at 56 GHz in bulk GaAs. In our experiment, we only see a moderate, smaller than a factor of two, increase in the lifetime. Consequently, the lifetime measured at 79 K must be mostly determined by extrinsic processes. The lifetime measured at RT contains both intrinsic and extrinsic contributions. Applying Matthiesen's rule[28], $\tau_{RT}^{-1} = \tau_{in}^{-1} + \tau_{ex}^{-1}$, we can estimate the intrinsic lifetime at RT as follows, $\tau_{in}^{-1} = \tau_{RT}^{-1} - \tau_{80K}^{-1}$, which yields $\tau_{in}$= 950 ps with the statistical error +/-74 ps.

According to Ref. 9, intrinsic phonon lifetime in GaAs-AlAs SLs is expected to be similar to that in bulk GaAs. First-principles calculations[12] of three-phonon scattering cross sections in GaAs indicated a quadratic dependence of RT phonon lifetime on frequency below ~ 2 THz. Extrapolating the calculated lifetimes[27] to 330 GHz yields the bulk longitudinal phonon lifetime of ~4 ns, i.e. significantly larger than found in our experiment. However, the three-phonon scattering model is only valid in the limit $\omega\tau_{th}$>>1, where $\tau_{th}$ is the lifetime of dominant thermal phonons.[15] In GaAs at RT, the thermal phonon lifetime can be as short as ~3ps[27]; therefore at 300 GHz this condition is no longer valid. In the opposite limiting case, $\omega\tau_{th}$<<1, phonon dissipation occurs in the Akhiezer relaxation regime.[15] The highest frequency for which the phonon relaxation in bulk GaAs has been measured[14] is 56 GHz. Ref. 14 reported a RT lifetime of 2.8 ns for longitudinal phonons and suggested that the measurement was within the range of the Akhiezer model.

Figure 4 shows experimental data from Ref. 14 and ultrasonic measurements[29,30] fitted by the Akhiezer model. Also shown is the theoretical three-phonon scattering trend [27] yielding a $1/\omega^2$ frequency dependence. At ultrasonic frequencies, Akhiezer model also yields a quadratic frequency dependence, albeit with a factor almost three orders of magnitude smaller compared to the Landau-Rumer model. In the high-frequency limit of the Akhiezer model, the lifetime is independent on frequency.[15] One would expect a transition between Akhiezer and Landau-Rumer regimes somewhere in the sub-THz range.[14] Figure 4 indicates that our measurement, shown by the filled circle, is likely to lie in this transition region.

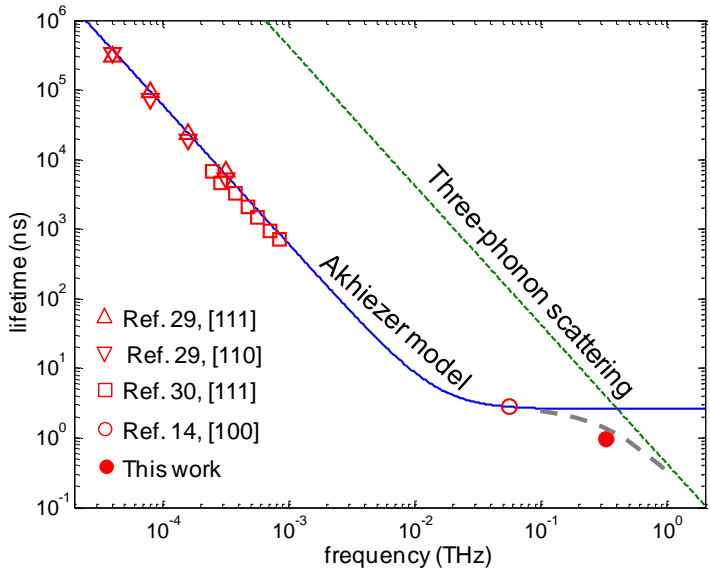

Fig. 4. Intrinsic longitudinal phonon lifetime at RT vs. frequency: literature data for bulk GaAs (open symbols) fitted by the Akhiezer relaxation model and extrapolated first-principles calculations for bulk GaAs with three-phonon scattering (Landau-Rumer) model[27] compared with the SL measurement from this work (solid circle). Dashed grey curve indicates a possible transition between the Akhiezer and Landau-Rumer regimes.

The only other measurement of an intrinsic phonon lifetime in a similar material at a comparable frequency was done on bulk GaN at 380 GHz,[31] with the reported lifetime of 150 ps. The authors suggested that the phonon relaxation was within the range of validity of the three-phonon scattering model. A simple estimate based on thermal conductivity values indicates that phonon lifetimes in GaN are expected to be comparable or larger than those in GaAs. The fact that the phonon lifetime in bulk GaN is much smaller compared to a GaAs-AlAs SL at a comparable frequency is unexpected and calls for a further investigation.

In summary, femtosecond pump-probe technique has been used to measure lifetime of the longitudinal zone-center mode of a 16-nm period SL at ~340 GHz. We conclude that the phonon lifetime measured at 79 K is dominated by extrinsic processes while at RT the extrinsic and intrinsic contributions to the lifetime are about equal. A comparison of the measured intrinsic lifetime of 0.95 ns with available theoretical and experimental results for bulk GaAs indicates that the observed phonon decay corresponds to the transition region between three-phonon scattering and Akhiezer relaxation models. We hope that our measurements will stimulate the interest to this transition and that more comprehensive experimental data and theoretical analysis for both bulk and nanostructured materials will become available soon.

The authors would like to thank Kara Manke and Jeff Eliason for help in the experiment and Jivtesh Garg for discussions of theoretical aspects. This work was supported as part of the S3TEC Energy Frontier Research Center funded by the U.S. Department of Energy, Office of Science, Office of Basic Energy Sciences under Award No. DE-SC0001299.

**References**

[1] D. G. Cahill, W. K. Ford, K. E. Goodson, G. D. Mahan, H. J. Maris, A. Majumdar, R. Merlin and S. R. Phillpot, J. Appl. Phys. **93,** 793 (2003).

[2] A.A. Balandin, J. Nanosci. Nanotechnol. **5**, 1015 (2005).

[3] M. S. Dresselhaus, G. Chen, M.Y. Tang, R.G. Yang, H. Lee, D.Z. Wang, Z.F. Ren, J.-P. Fleurial, and P. Gogna, Adv. Mater. **19**, 1043–1053 (2007).

[4] E. Pop, Nano Res. **3**, 147 (2010).

[5] S. Tamura, D. C. Hurley and J. P. Wolfe, Phys. Rev. B **38**, 1427 (1987).

[6] A. Yamamoto, T. Mishina, and Y. Masumoto, Phys. Rev. Lett. **73**, 740 (1994).

[7] M. Trigo, T. A. Eckhause, M. Reason, R. S. Goldmanand, and R. Merlin, Phys. Rev. Lett. **97**, 124301 (2006).

[8] A. Huynh, B. Perrin, B. Jusserand, and A. Lemaître, Appl. Phys. Lett. **99**, 191908 (2011).

[9] A. Ward and D. A. Broido, Phys. Rev. B **77**, 245328 (2008).

[10] A. Ward and D. A. Broido, Phys. Rev. B **81**, 085205 (2010).

[11] K. Esfarjani, G. Chen and H.T. Stokes, Phys. Rev. B **84**, 085204 (2011).

[12] T. Luo, J. Garg, J. Shiomi, K. Esfarjani, and G. Chen, Europhys. Lett., submitted.

[13] B. C. Daly, K. Kang, Y. Wang, and D. G. Cahill, Phys. Rev. B **80**, 174112 (2009).

[14] W. Chen, H. J. Maris, Z. R. Wasilewski and S. Tamura, Phil. Mag. B **70**, 687 (1994).

[15] H. J. Maris, in *Physical Acoustics*, edited by W. P. Mason and R. N. Thurston (Academic, 1971), Vol. 8, p. 279.

[16] M. Trigo, *Ultrafast Dynamics of Folded Acoustic Phonons from Semiconductor Superlattices*, Ph.D. Thesis (University of Michigan, 2008).

[17] G. Rozas, M. F. Pascual Winter, B. Jusserand, A. Fainstein, B. Perrin, E. Semenova, and A. Lemaitre, Phys. Rev. Lett. **102**, 015502 (2009).

[18] P. Babilotte, P. Ruello, T. Pezeril, G. Vaudel, D. Mounier, J.-M. Breteau, and V. Gusev, J. Appl. Phys. **109**, 064909 (2011).

[19] V. E. Gusev and A. A. Karabutov, *Laser Optoacoustics* (American Institute of Physics, 1993).

[20] M. Grimsditch, R. Bhadra, I. K. Schuller, F. Chambers and G. Devane, Phys. Rev. B **42**, 5 (1990).

[21] M. R. Brozel and G. E. Stillmann, *Properties of Gallium Arsenide,* (INSPEC, 1996).

[22] R. B. Blackman, J. W. Tukey, "Particular Pairs of Windows." In *The Measurement of Power Spectra, From the Point of View of Communications Engineering*, (Dover, New York 1959), p. 98.

[23] M. F. Pascual-Winter, A. Fainstein, B. Jusserand, B. Perrin, and A. Lemaître, Chinese J. Phys. 49, 250 (2011).

[24] G. Bastard, *Wave Mechanics Applied to Semiconductor Heterostructures* (Les Editions de Physique, 1988).

[25] S. Adachi, *GaAs and Related Materials: Bulk Semiconducting and Superlattice Properties* (World Scientific, 1994).

[26] C. Herring, Phys. Rev. **95**, 954 (1954).

[27] Calculated intrinsic phonon lifetimes in bulk GaAs data were provided by the authors of Ref. 12.

[28] J.M. Ziman, *Electrons and Phonons* (Oxford University Press, 2001).

[29] R. I. Cottam and G. I. Saunders, J. Phys. C **7**, 2447 (1974).

[30] S. Ivanov, J.M. Kotelyanskii, G.D. Mansfeld, and E.N. Khazanov, Sov. Phys. Solid State **13**, 508 (1971).

[31] T.-M. Liu, S.-Z. Sun, C.-F. Chang, C.-C. Pan, G.-T. Chen, J.-I. Chyi, V. Gusev, and C.-K. Sun, Appl. Phys. Lett. **90**, 041902 (2007).